\documentclass[amsmath, amssymb, reprint, aip]{revtex4-2}
\pdfoutput=1

\usepackage{graphicx} 
\usepackage{hyperref} 
\usepackage{mathptmx} 
\usepackage{xcolor}   
\usepackage{dcolumn}  
\usepackage{ulem}
\usepackage{orcidlink}

\DeclareMathAlphabet{\pazocal}{OMS}{zplm}{m}{n} 

\hypersetup{
    colorlinks=true,
    linkcolor=blue,
    citecolor=blue,
    filecolor=blue,
    urlcolor=blue
}

\graphicspath{{./}{./figs/}}

\newcommand{\SSD}{Sensor Science Division, National Institute of Standards and Technology, Gaithersburg, Maryland 20899, USA}
\newcommand{\QMD}{Quantum Measurement Division, National Institute of Standards and Technology, Gaithersburg, Maryland 20899, USA}
\newcommand{\JQI}{Joint Quantum Institute, College Park, Maryland 20742, USA}
\newcommand{\UMD}{Physics Department, University of Maryland, College Park, Maryland, 20742, USA}

\newcolumntype{d}{D{.}{.}{4.4}}

\begin{document}

\title{On the effect of ``glancing'' collisions in the cold atom vacuum standard}

\author{Stephen P. Eckel\,\orcidlink{0000-0002-8887-0320}}
\email{stephen.eckel@nist.gov}
\affiliation{\SSD}
\author{Daniel S. Barker\,\orcidlink{0000-0002-4614-5833}}
\affiliation{\SSD}
\author{James A. Fedchak\,\orcidlink{0000-0002-1259-0277}}
\affiliation{\SSD}
\author{Jacek K{\l}os\,\orcidlink{0000-0002-7407-303X}}
\affiliation{\JQI}\affiliation{\UMD}
\author{Julia Scherschligt\,\orcidlink{0000-0003-4965-0103}}
\affiliation{\SSD}
\author{Eite Tiesinga\,\orcidlink{0000-0003-0192-5585}}
\affiliation{\JQI}\affiliation{\UMD}\affiliation{\QMD}

\date{\today}

\begin{abstract}
We theoretically investigate the effect of ``glancing" collisions on the ultra-high vacuum (UHV) pressure readings of the cold atom vacuum standard (CAVS), based on either ultracold $^7$Li or $^{87}$Rb atoms.
Here, glancing collisions are those collisions between ultracold atoms and room-temperature background atoms or molecules in the vacuum that do not impart enough kinetic energy to eject an ultracold atom from its trap.
Our model is wholly probabilistic and shows that the number of the ultracold atoms remaining in the trap as a function of time is non-exponential. 
We update the recent results of a comparison between a traditional pressure standard---a combined flowmeter and dynamic expansion system---to the CAVS [D.S. Barker, {\it et al.}, {\it AVS Quantum Science} {\bf 5} 035001 (2023)] to reflect the results of our model.
We find that the effect of glancing collisions shifts the theoretical predictions of the total loss rate coefficients for $^7$Li colliding with noble gases or N$_2$ by up to $0.6$~\%.
Likewise, we find that in the limit of  zero trap depth the experimentally extracted  loss rate coefficients for $^{87}$Rb colliding with noble gases or N$_2$ shift by as much as 2.2~\%.
\end{abstract}

\maketitle

\section{Introduction}
\label{sec:intro}
Cold atom vacuum standards measure vacuum pressures by observing the loss rate of cold, ``sensor'' atoms from a shallow, conservative trap due to collisions with background gases at temperature $T$ and number density $n$.\cite{Bjorkholm1988, Fagnan2009, US8803072, Arpornthip2012, Yuan2013, Moore2015, Makhalov2016, Scherschligt2017, Eckel2018, Booth2019, Shen2020, Shen2021, Barker2022, Klos2023, Barker2023}
Most collisions impart sufficient energy $\Delta E$ such that the  total energy $E$ of the sensor atom after the collision is larger than the trap depth $W$, causing a sensor atom to be ejected from the trap and lost.
If all collisions result in loss, then the sensor atom loss is exponential in time with loss rate  $\Gamma = K n$, where $K$ is the collision rate coefficient with SI unit cm$^3$/s and $n$ with SI unit cm$^{-3}$.
However, some collisions are ``glancing'': they do not impart enough energy to eject an atom from the trap.
These glancing collisions represent an important systematic effect for CAVS-based vacuum pressure measurements.

How to treat systematics due to the glancing collisions has been a matter of discussion in the literature.
The simplest means to assess systematics is to assume, as done implicitly in Refs.~\onlinecite{Eckel2018, Booth2019, Makrides2019, Makrides2020, Shen2020, Makrides2022Ea, Makrides2022Eb, Klos2023, Barker2023},
(1) that the temperature of the sensor atom cloud $T_{\rm S}$ is zero, and (2) that glancing collisions do not change the energy of a sensor atom. 
Under these two assumptions, sensor atom loss is still exponential but now with rate $\Gamma = K n [1-P_{\rm A}(W)]$, where $P_{\rm A}(\Delta E)$ is the cumulative probability that a collision adds a positive energy less than $\Delta E$ to the sensor atom's kinetic energy in the trap.
(See App.~\ref{sec:app:kinematics} for an explanation of the kinematics and the sign of $\Delta E$.)
Various theoretical expressions for $P_{\rm A}(W)$ exist in the literature. 
For example, Ref.~\onlinecite{Klos2023} gives $P_{\rm A}(W)=(a_{\rm gl} W + b_{\rm gl} W^2)/K$ by appropriately factoring Eq.~1 in their paper.
Other expressions can be found in Refs.~\onlinecite{Eckel2018,Booth2019, Shen2020, Shen2021, Stewart2022, Ehinger2022, Barker2022, Shen2023}.

Estimates show that if $^{87}$Rb is used as a sensor atom, $P_{\rm A}(W)$ is of the order of $10$~\% for most collision partners at $T=300$~K and $W/k_{\rm B}\sim 1$~mK, where $k_{\rm B}$ is the Boltzmann constant.
For $^7$Li sensor atoms, $P_{\rm A}(W)$ is about an order of magnitude smaller.
Reference~\onlinecite{Shen2021} addressed the first of these two assumptions and derived a modified form for the observed loss rate for $T_{\rm S}>0$.
Reference~\onlinecite{Stewart2022} relaxed the second of these two assumptions in the context of Rb+Rb collisions, numerically calculating a glancing-collision corrected measurement of the $C_6$ van-der-Waals  coefficient assuming the universality of quantum diffractive collisions (UQDC).
Likewise, Ref.~\onlinecite{Deshmukh2024} relaxed both assumptions and found that the measured collision rates could be shifted by as much as 3~\% for Ar background atoms colliding with $^{87}$Rb.

A recent comparison by the current authors between cold atom vacuum standards and a classical vacuum metrology apparatus showed uncertainties approaching 1~\%,\cite{Barker2023} but used the two simplistic assumptions described above to fit for $K$ and the first two terms of the Taylor expansion of $P_{\rm A}(\Delta E)$ in terms of $\Delta E$.
At the 1~\% level of accuracy, the effect of two glancing collisions per sensor atom within a time $t\lesssim 3/\Gamma$ is important for $^{87}$Rb, because $[P_{\rm A}(W)]^2\sim 1$~\%.
This glancing collision effect was not included in the uncertainty budget of Ref.~\onlinecite{Barker2023}.

Inspired by the developments of Refs.~\onlinecite{Shen2021, Stewart2022, Deshmukh2024}, we strive to create a comprehensive and fully analytic model of glancing collisions in the CAVS.
We structure our model, described in Sec.~\ref{sec:analysis}, in terms of two probabilities: the probability for a sensor atom to have a collision with a background gas atom or molecule and the probability of that collision to cause the sensor atom to acquire kinetic energy $E>W$; causing that atom to be ejected from the trap.
We show that under certain conditions, the loss of sensor atoms from the trap becomes non-exponential.
In Sec.~\ref{sec:effect}, we correct the recent data of Ref.~\onlinecite{Barker2023} based on our analytic model.
In App.~\ref{sec:app:kinematics}, we give a brief review of the kinematics, while in App.~\ref{sec:app:temperature}, we refine the estimate of the temperature of the $^{87}$Rb cloud used in Ref.~\onlinecite{Barker2023}.

\section{Probabilistic Model of Sensor-Atom Loss}
\label{sec:analysis}

Our probabilistic model of atom loss begins by noting that the ultracold sensor atom
clouds are dilute and far away from quantum degeneracy, i.e., the average time between collisions among sensor atoms is much longer than $1/(Kn)$ and thus the sensor-atom cloud  is effectively a non-interacting gas of atoms.
For the $^{7}$Li gases in Ref.~\onlinecite{Barker2023}, the rate of collisions between sensor atoms $\beta \approx K_{\rm S}n_{\rm S}\sim10^{-6}$ s$^{-1}$ at the maximum possible number densities of $n_{\rm S}\sim 10^6$ cm$^{-3}$ and  rate coefficient $K_{\rm S}\approx 1.5\times10^{-12}$~cm$^3$/s.
This rate is much smaller than any measured $\Gamma$ in the UHV domain by Ref.~\onlinecite{Barker2023}.
For the $^{87}$Rb gases in Ref.~\onlinecite{Barker2023}, $\beta\approx 0.03$~s$^{-1}$, when averaged over the entire, non-uniform density of the cloud using a peak density of $n_{\rm S}\approx 1.6\times10^{-9}$~cm$^{-3}$, a temperature $T_{\rm S}=53(11)$~$\mu$K (see App.~\ref{sec:app:temperature}), and $K_{\rm S}\sim 1.2\times 10^{-10}$ cm$^3$/s.
This rate is a factor of two smaller than the smallest measured $\Gamma$ in Ref.~\onlinecite{Barker2023}.
We thus need only to construct a probabilistic loss model for an ensemble of traps, each containing a 
single sensor atom. An atom has one or more collisions with atoms or molecules in the vacuum before 
it leaves its trap.

The experimental design of traps for ultracold atoms has a significant impact on the equilibrium or
initial $t=0$ probability distribution of the atoms. First, as mentioned before,
these traps have a finite depth $W$ but, in addition, in the process
of cooling the sensor atoms the experimentalists use what is informally called a ``knife'' that
removes any atom with a motional energy larger than cut-off energy $E_{\rm c}$ that is often 
significantly lower than $W$. Moreover, the trapping potential $U(\mathbf{r})$
as function of location $\mathbf{r}$ limits the spatial excursions of the atoms.
Assuming that the potential energy of the trap is smallest at $\mathbf{r}=\mathbf{0}$ and that $U(\mathbf{0})=0$, the cumulative probability of a sensor atom having an energy less than $E$ is given by
\begin{equation}
    P_0(E) = \int_0^E {\rm d}E'\rho(E') e^{-E'/k_{\rm B}T_{\rm S}} \,,\label{eq:P0_fundamental}
\end{equation}
where $\rho(E)$ is the density of states and depends on the shape of $U(\mathbf{r})$.
For a separable, power law potential of the form
\begin{equation}
    U(x,y,z) = c_1|x|^p + c_2 |y|^q + c_3 |z|^s\,,
\end{equation}
where $p,q,s>0$, $|x|$ is the absolute value of real $x$, and $c_i>0$ for $i=1$, 2, and 3, the density of states obeys~\cite{Bagnato1987, Luiten1996, pethick2002bose}
\begin{equation}
    \rho(E) \propto E^{1/2+1/p+1/q+1/r} \equiv E^{1/2+\delta}\ .
\end{equation}
Equation~\eqref{eq:P0_fundamental} can then be evaluated analytically and is
\begin{eqnarray}
    P_0(T_{\rm S}, E_{\rm c},E) & = &
    \frac{\gamma(3/2+\delta, E/k_{\rm B} T_{\rm S})}{\gamma(3/2+\delta, E_{\rm c}/k_{\rm B} T_{\rm S})} 
    \label{eq:PI}
\end{eqnarray}
when $E<E_{\rm c}$ and 1 otherwise.
Here, $\gamma(a,x) = \int_0^x y^{a-1} e^{-y}\ dy$ is the incomplete gamma function. 

For a magneto-optical trap (MOT) with its linear trapping force and thus quadratic potential along all three spatial dimensions, $\delta=3/2$. 
For a magnetic quadrupole trap with constant trapping force and thus a linear potential along all three spatial axes, $\delta=3$.
For a constant or box potential, $\delta=0$.
The sensor-atom clouds created in Ref.~\onlinecite{Barker2023} were last in thermal equilibrium in a MOT, and thus we anticipate that $\delta=3/2$.
Subsequent measurements, described in App.~\ref{sec:app:temperature}, confirm this.

For time $t>0$, we observe that the cumulative probability $P_{\rm R}(E,t)$ for an atom in the ensemble to have total energy less than $E$ can be expressed in terms of the expansion
\begin{eqnarray}
    P_{\rm R}(E,t) & = & p_{0}^{(c)}(t)P_0( T_{\rm S}, E_{\rm c},E) + p_{1}^{(c)}(t) P_1( T_{\rm S}, E_{\rm c},E) \nonumber \\
    & & \quad +\, p_{2}^{(c)}(t) P_2( T_{\rm S}, E_{\rm c},E) + \cdots \,,
    \label{eq:prob}
\end{eqnarray}
where $p_{k}^{(c)}(t)$ is the probability that a sensor atom  has experienced $k=0$, 1, 2, $\dots$  collisions with background gas atoms or molecules after time $t$. 
Collisions between a sensor atom and background atoms and molecules occur at random times with a timescale of $1/(Kn)$.
The distribution of collision times is then Poissonian and 
\begin{equation}
    \label{eq:poisson}
    p_{k}^{(c)}(t) = \frac{1}{\Gamma(k+1)}(K n t)^k e^{-K n t}\,,
\end{equation}
where $\Gamma(z)$ is the Gamma function.

Cumulative probability $P_0(E, T_{\rm S}, E_{\rm c})$ is defined in Eq.~(\ref{eq:PI}) and, more generally, $P_k(E, T_{\rm S}, E_{\rm c})$ are the cumulative probabilities that an atom has total energy less than $E$ after $k=0$, 1, 2, $\dots$ collisions.
We can derive $P_k(E, T_{\rm S}, E_{\rm c})$ recursively.
After the $k$-th collision,  $P_k(E, T_{\rm S}, E_{\rm c})$ is given by the convolution
of the probability density $p_{\rm A}(\Delta E)$ that the collision adds  energy $\Delta E>0$
to the sensor atom and the cumulative probability $P_{k-1}(E', T_{\rm S}, E_{\rm c})$
such that $E=E'+\Delta E$. Following App.~\ref{sec:app:kinematics} we realize that nearly all
collisions add energy to the sensor atoms and assume $p_{\rm A}(\Delta E)=0$ for $\Delta E<0$. Consequently,
\begin{equation}
    \label{eq:P_i}
    P_k( T_{\rm S}, E_{\rm c},E) = \int_0^\infty {\rm d}E' p_{\rm A}(E-E')P_{k-1}( T_{\rm S}, E_{\rm c},E')\,.
\end{equation}
The probability density $p_{\rm A}(\Delta E)$  is found by noting that 
\begin{equation}
    \int_0^{\Delta E} {\rm d}\epsilon p_{\rm A}(\epsilon)  \equiv P_{\rm A}(\Delta E)
    \label{eq:littlep}
\end{equation}
with cumulative probability $P_{\rm A}(\Delta E)$ as defined in the introduction.
It reasonable to assume that $p_{\rm A}(\Delta E)$ and $P_{\rm A}(\Delta E)$ are independent of the initial collision energy as long as $\Delta E\gg k_{\rm B}T_{\rm S}$, with sensor atoms essentially at rest, and index $k$ limited to only a few collisions.
As discussed in Refs.~\onlinecite{Makrides2019, Makrides2020, Barker2023}, the change in $P_{\rm A}(\Delta E)$ due to a finite initial velocity of the ultracold atoms is of the order of $(m/\mu)(T_{\rm S}/T) \sim 10^{-5}$, where $\mu$ is the reduced mass of the collision partners, and is  negligible at our level of accuracy.

\begin{table}
    \begin{tabular}{c|dd}
    \hline\hline
       System & \multicolumn{1}{c}{$\alpha_1/k_{\rm B}$} & \multicolumn{1}{c}{$\alpha_2/k^2_{\rm B}$}\\
         & \multicolumn{1}{c}{(mK$^{-1}$)} & \multicolumn{1}{c}{(mK$^{-2}$)} \\
       \hline
       $^{87}$Rb-H$_2$ & 0.037 & -0.002 \\
       $^{87}$Rb-He & 0.014 & -0.00028 \\
       $^{87}$Rb-Ne & 0.053 & -0.003 \\
       $^{87}$Rb-N$_2$ & 0.076 & 0.0068 \\ 
       $^{87}$Rb-Ar & 0.079 & -0.0072 \\
       $^{87}$Rb-Kr & 0.11 & -0.014 \\
       $^{87}$Rb-Xe & 0.14 & -0.024 \\       
       \hline\hline
    \end{tabular}
    \caption{Values of coefficients $\alpha_i$ for $^{87}$Rb sensor atoms with $i=1$ and $2$ in Eq.~\eqref{eq:P_I} derived from the theory of Ref.~\onlinecite{Klos2023}. No uncertainties are presented as the table is only meant to indicate relative sizes of the $\alpha_i$. Values for $\alpha_i$ with $i>2$ are currently unknown and assumed to be zero in our simulations.}
    \label{tab:alphas}
\end{table}

From the introduction, we also have
\begin{equation}
    P_{\rm A}(\Delta E) = \alpha_1 \Delta E + \alpha_2 \times (\Delta E)^2 + \cdots 
    \label{eq:P_I}
\end{equation}
for $\Delta E\ge 0$ with $\alpha_1=a_{\rm bg}/K$ and $\alpha_2=b_{\rm bg}/K$.  Consequently,
\begin{equation}
    p_{\rm A}(\Delta E) =\sum_{i=1}^\infty i(\Delta E)^{i-1} \alpha_i
    \label{eq:p_I}
\end{equation}
for $\Delta E\ge 0$ and zero otherwise.
We have used $K$, $a_{\rm gl}$, and $b_{\rm gl}$ from Ref.~\onlinecite{Klos2023},
reproduced in Table \ref{tab:Rb_results} for $^{87}$Rb,
and realize that the $P_{\rm A}(\Delta E)$ are much smaller than one for the relevant $\Delta E$.
Table~\ref{tab:alphas} lists values for $^{87}$Rb sensor atoms. Typical trap depths are of order $k_{\rm B}\times 1$ mK and we realize that $\alpha_2W^2\ll \alpha_1 W\ll 1$.
For $^7$Li sensor atoms these inequalities also hold.
An alternative choice from Refs.~\onlinecite{Booth2019, Shen2020, Shen2021, Stewart2022, Shen2023} results in $\alpha_j=\beta_j/(U_d)^j$, where $U_d=4\pi \hbar^2/[m (K/\langle v\rangle)]$ is the median energy exchanged in the collision with mean velocity $\langle v\rangle=\sqrt{k_{\rm B} T/2 M}$ for a background gas species with mass $M$. Here, $\hbar$ is the reduced Planck constant.
Coefficients $\beta_j$ are given in Table~1 of Ref.~\onlinecite{Booth2019}.

The probabilities $P_i(T_{\rm S}, E_{\rm c},E)$ can be evaluated using 
the chapter on the confluent hypergeometric function found in Ref.~\onlinecite{NIST:DLMF}
after we  express the incomplete gamma function in terms of the Kummer's confluent hypergeometric function.
After some thought, we obtain 
\begin{widetext} 
\begin{eqnarray}
P_1(T_{\rm S}, E_{\rm c}, E) & = & \sum_{i=1}^\infty  \alpha_i (k_{\rm B}T_{\rm S})^i
   \left\{ \begin{array}{cl}
      \displaystyle        {\cal M}_i\left(\frac{3}{2}+\delta,\frac{E_{\rm c}}{k_{\rm B}T_{\rm S}},\frac{E}{k_{\rm B}T_{\rm S}}\right) & E<E_{\rm c}\\
    \displaystyle  \sum_{k=0}^{i}
          \left({i\atop k}\right)  \left(\frac{E-E_{\rm c}}{k_{\rm B}T_{\rm S}}\right)^k {\cal M}_{i-k}\left(\frac{3}{2}+\delta,\frac{E_{\rm c}}{k_{\rm B}T_{\rm S}},\frac{E_{\rm c}}{k_{\rm B}T_{\rm S}}\right) 
           & E\ge E_{\rm c}
          \end{array}
          \right. \,,
           \label{eq:P1_full}
\end{eqnarray}
\end{widetext}
where $ \left({i\atop k}\right)$ is the binomial coefficient,
 dimensionless function
\begin{eqnarray}
 {\cal M}_i(a,x_{\rm c},x) &\equiv& i\int_0^{x} {\rm d}\epsilon (x-\epsilon)^{i-1} 
  \frac{\gamma(a, \epsilon)}{\gamma(a, x_{\rm c})} \label{eq:Mi} \\
    &=&  x^i \left(\frac{x}{x_{\rm c}}\right)^a
   \frac{\Gamma(i+1) \mathbf{M}(a,a+i+1,-x)}{\mathbf{M}(a,a+1,-x_{\rm c})}\,,
   \label{eq:Mi2}
 \end{eqnarray}
 and Kummer's (regularized) confluent hypergeometric function $\mathbf{M}(a,b,z)$ is defined in Eqs.~(\href{http://dlmf.nist.gov/13.2.E3}{13.2.E3}) and (\href{http://dlmf.nist.gov/13.2.E4}{13.2.E4}) of Ref.~\onlinecite{NIST:DLMF}.
In the derivation of Eq.~(\ref{eq:Mi2}), we have also used Eqs.~(\href{https://dlmf.nist.gov/13.6.E5}{13.6.E5}) and (\href{http://dlmf.nist.gov/13.4.E2}{13.4.E2}) of this reference.
We observe that Eq.~(\ref{eq:Mi2}) allows us to define ${\cal M}_i(a,x_{\rm c},x)$ for $i=0$
and realize that ${\cal M}_0(a,x_{\rm c},x_{\rm c})=1$.
Finally, note that  
\begin{equation}
  {\cal M}_i(a,x_{\rm c},x_{\rm c}) \to  x_{\rm c}^i  
  \left( 1- i\frac{a}{x_{\rm c}} + O(1/x_{\rm c}^2)\right)
 \end{equation}
for $x_{\rm c}\to+\infty$ based on Eq.~(\href{http://dlmf.nist.gov/13.7.E2}{13.7.E2}) of Ref.~\onlinecite{NIST:DLMF}.
Consequently, focusing on $E\approx W$ and realizing that $k_{\rm B} T_{\rm S}\ll E_{\rm c} < W$,  we have
\begin{eqnarray}
\lefteqn{   P_1(T_{\rm S},E_{\rm c},E) \to \sum_{i=1}^\infty \alpha_i \left[\sum_{k=0}^{i}
    \left({ i\atop k }\right)   (E-E_{\rm c})^k  E_{\rm c}^{i-k} \right.
  }  \\
    && \left.- (3/2+\delta) k_{\rm B}T_{\rm S}    
          \sum_{k=0}^{i}
    \left({ i\atop k }\right) (i-k)  (E-E_{\rm c})^k  E_{\rm c}^{i-k-1} 
    \right]\nonumber \\
    &= & \sum_{i=1}^\infty \alpha_i E^i\left[1 - (3/2+\delta) i \frac{k_{\rm B}T_{\rm S}}{E} \right]
    \label{eq:P1power}\\
    &=&P_{\rm A}(E) -(3/2+\delta) k_{\rm B}T_{\rm S} p_{\rm A}(E) \,,
    \label{eq:P:11power}
\end{eqnarray}
where in the formula in square brackets we recognize the binomial formula and its derivative. 

Next, we derive
\begin{widetext}
\begin{eqnarray} 
   P_2(T_{\rm S},E_{\rm c},E)&=&\sum_{j=1}^\infty\sum_{i=1}^\infty \alpha_j\alpha_i
   (k_{\rm B} T_{\rm S})^{i+j}
   \frac{\Gamma(i+1)\Gamma(j+1)}{\Gamma(i+j+1)}
  \left\{ \begin{array}{cl}
  \displaystyle
  {\cal M}_{i+j}\left(\frac{3}{2}+\delta,\frac{E_{\rm c}}{k_{\rm B}T_{\rm S}},\frac{E}{k_{\rm B}T_{\rm S}}\right)  
     & E<E_{\rm c}\\
  \displaystyle  \sum_{k=0}^{i+j} \left({i+j\atop k}\right) \left(\frac{E-E_{\rm c}}{k_{\rm B}T_{\rm S}}\right)^{k}  
   {\cal M}_{i+j-k}\left(\frac{3}{2}+\delta,\frac{E_{\rm c}}{k_{\rm B}T_{\rm S}},\frac{E_{\rm c}}{k_{\rm B}T_{\rm S}}\right)       & E\ge E_{\rm c}
     \end{array}
  \right.
  \label{eq:P2_full}
\end{eqnarray}
again using Eq.~(\href{http://dlmf.nist.gov/13.4.E2}{13.4.E2}) of Ref.~\onlinecite{NIST:DLMF}. 
Again focusing on $E\approx W$ and realizing that $k_{\rm B} T_{\rm S}\ll E_{\rm c} < W$, we have 
 \begin{equation}
    P_2(T_{\rm S},E_{\rm c},E)\to   
     \sum_{j=1}^\infty\sum_{i=1}^\infty \alpha_j\alpha_i\frac{\Gamma(i+1)\Gamma(j+1)}{\Gamma(i+j+1)} E^{i+j} 
    = \frac{1}{2} \alpha_1^2E^2 + \frac{2}{3} \alpha_1\alpha_2 E^3 + \cdots
   \label{eq:P2power}
\end{equation}
as we again recognized a binomial formula in $E-E_{\rm c}$ and $E_{\rm c}$.
We also observe that $P_2(T_{\rm S},E_{\rm c},E) \approx [P_{\rm A}(E)]^2/2$.
 
For completeness, we have
\begin{eqnarray}
    \label{eq:P3_full}
  \lefteqn{ P_3(T_{\rm S},E_{\rm c},E)=
  \sum_{l=1}^\infty\sum_{j=1}^\infty\sum_{i=1}^\infty \alpha_l\alpha_j\alpha_i (k_{\rm B}T_{\rm S})^{i+j+l} } \\
    &&\quad \quad\quad\quad\quad \times   \frac{\Gamma(i+1)\Gamma(j+1)\Gamma(l+1)}{\Gamma(i+j+l+1)} 
    \left\{ \begin{array}{cl}
      \displaystyle
  {\cal M}_{i+j+l}\left(\frac{3}{2}+\delta,\frac{E_{\rm c}}{k_{\rm B}T_{\rm S}},\frac{E}{k_{\rm B}T_{\rm S}}\right)  
     & E<E_{\rm c}\\
  \displaystyle  \sum_{k=0}^{i+j+l} \left({i+j+l\atop k}\right) \left(\frac{E-E_{\rm c}}{k_{\rm B}T_{\rm S}}\right)^{k} 
   {\cal M}_{i+j+l-k}\left(\frac{3}{2}+\delta,\frac{E_{\rm c}}{k_{\rm B}T_{\rm S}},\frac{E_{\rm c}}{k_{\rm B}T_{\rm S}}\right)       & E\ge E_{\rm c}
       \end{array}
       \right.
       \nonumber 
    \end{eqnarray}
\end{widetext}
The expressions for $P_k(T_{\rm S},E_{\rm c},E)$ for $k=4, 5, \cdots$ follow by inspection from Eqs.~(\ref{eq:P1_full}), (\ref{eq:P2_full}), and (\ref{eq:P3_full}).
For $k_{\rm B}T_{\rm S}\ll E_{\rm c}< E$ we then  derive
 \begin{eqnarray}
 \lefteqn{  P_k(T_{\rm S},E_{\rm c},E) \approx  \frac{1}{\Gamma(k+1)} \alpha_1^k E^k
   \left( 1 - k(3/2+\delta) \frac{k_{\rm B}T}{E}
   \right)}
   \label{eq:PkAsymptotic} \\
    &&
     + \frac{2k}{\Gamma(k+2)} \alpha_1^{k-1}\alpha_2 E^{k+1}
      \left( 1 - (k+1)(3/2+\delta) \frac{k_{\rm B}T}{E}
   \right)
    \nonumber
 \end{eqnarray}
 for $k=0$, 1, $\dots$ leading to an approximate analytical expression for the time evolution of the cumulative probability for an 
 atom in the ensemble to have total energy less than  $E$ that is given by
 \begin{eqnarray}
   P_{\rm R}^{\rm approx}(E,t) 
   &=&
      e^{-K n t} \biggl\{ I_0(2y) 
      - (3/2+\delta) \frac{k_{\rm B}T}{E}  y\, I_1(2y) \nonumber\\
      && \quad  + 2\frac{\alpha_2 E}{\alpha_1}
      \left( 1 - 2(3/2+\delta) \frac{k_{\rm B}T}{E} \right)  I_2(2y)
      \nonumber \\
          && \qquad
          \left.
          - 2(3/2+\delta) \frac{\alpha_2 k_{\rm B}T}{\alpha_1} y\, I_3(2y)
      \right\}  \label{eq:BesselI}
 \end{eqnarray}
with $y=\sqrt{\alpha_1 E Kn t}$ and is 
thus non-exponential in time. Here, $I_n(z)$ is  the modified Bessel function of the first kind, $I_n(z)\to (z/2)^n/\Gamma(n+1)$ for $z\to0$, and $I_n(z)\to e^z/\sqrt{2\pi z}$ for $z\to\infty$.
Equations (\ref{eq:PkAsymptotic}) and (\ref{eq:BesselI}) form two of the main analytical results of this article.

Two further, less accurate approximations for the cumulative probability $P_{\rm R}(E,t)$ are relevant to compare with previous work. We can find these approximations by rewriting Eqs.~(\ref{eq:prob}) and (\ref{eq:poisson}) in the equivalent cumulant form
\begin{equation}
P_{\rm R}(E,t)=   P_0 e^{-(1-P_1/P_0)Knt+(P_0P_2-P_1^2)(Knt)^2/(2P_0^2)
            +O(t^3)}       \label{eq:thirdorder}
\end{equation}
suppressing the three arguments of the cumulative probabilities $P_k(T_{\rm S},E_{\rm c},E)$ for clarity.
The terms proportional to $t^2$, $t^3$, etc in the exponential lead to non-exponential behavior.
If only assumption (2) holds, {\it i.e.} glancing collisions do not change the energy of an atom,
then the cumulative probability that an atom survives $i$ uncorrelated collisions is equal to the probability that an atom survives one collision raised to the $i$th power.
This corresponds to the choice  $P_k(T_{\rm S},E_{\rm c},E) = [P_1(T_{\rm S},E_{\rm c},E)]^k$ for $k=0$,1, 2, $\dots$ and only  the term linear in $t$ in Eq.~(\ref{eq:thirdorder}) survives 
leading to
\begin{equation}
    P_{\rm R}^{(1)} (E,t) 
    = e^{-[1-P_1(T_{\rm S}, E_{\rm c},E)] K nt}\,, \label{eq:exponential}
\end{equation}
equivalent to the result of Ref.~\onlinecite{Shen2021}.
When both assumptions hold, we have $P_k(T_{\rm S},E_{\rm c},E) = [P_{\rm A}(E)]^k$ for $k=0$, 1, 2, $\dots$ and 
\begin{equation}
    P_{\rm R}^{({\rm A})} (E,t) 
    = e^{-[1-P_{\rm A}(E)] Knt} \label{eq:Barker}
\end{equation}
used in Refs.~\onlinecite{Shen2020, Klos2023, Barker2022, Barker2023}, respectively.
With Eq.~(\ref{eq:BesselI}), we have shown that the exponential time evolutions in Eqs.~(\ref{eq:exponential}) and (\ref{eq:Barker}) are only valid
up to first order in $P_{\rm A}(E)$ or, more precisely, up to first order in $\alpha_1E $.

Finally, we note that the fraction of remaining trapped sensor atoms in the ensemble $\eta_{\rm S}(t)$ is given by $\eta_{\rm S}(t) = P_{\rm R}(E=W,t)$.\footnote{Note that in Ref.~\onlinecite{Barker2023} for practical reasons, $\eta_{\rm S}(t)$ was the ratio of the measured atom number in the magnetic trap after time $t$ to the atom number in the magneto-optical trap (MOT) just before transfer into the magnetic trap.}
Inspection, remembering that $E_{\rm c}\leq W$, shows that $\eta_{\rm S}(t=0) =1$.
For future use, we define a measure of non-exponential behavior by
\begin{equation}
    \Xi(W,t)= \left|P_2(T_{\rm S},E_{\rm c},W)-[P_1(T_{\rm S},E_{\rm c},W)]^2\right| Knt
\end{equation}
based on the ratio of the $t^2$ and $t$ terms in Eq.~(\ref{eq:thirdorder}), where  we have used that $P_0(T_{\rm S},E_{\rm c},W)=1$ and $P_1(T_{\rm S},E_{\rm c},W)\ll1$ for our sensor atoms.

\begin{figure*}
    \centering
    \includegraphics{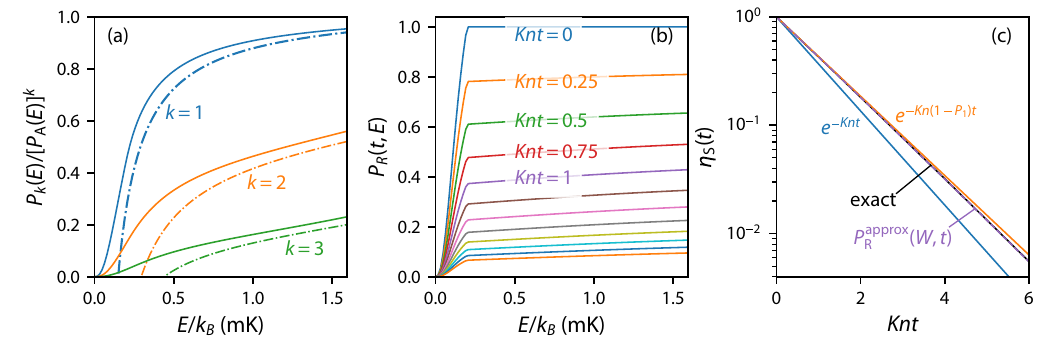}
    \caption{
    (a) Exact and approximate scaled cumulative probabilities $P_k(T_{\rm S},E_{\rm c},E)/[P_{\rm A}(E)]^k$ for $^{87}$Rb sensor atoms having an energy less than $E$ after $k=1,2,3$ collisions with Xe background gas atoms in blue, orange, and green, respectively. We use the  $P_{\rm A}(\Delta E)$ from Ref.~\onlinecite{Klos2023}
    and assume  $\delta=3/2$, $T_{\rm S}$=53~$\mu$K, and $E_{\rm c}/k_{\rm B}=0.2$~mK.
    Solid curves are based on the exact expressions in Eqs.~\eqref{eq:P1_full}, \eqref{eq:P2_full}, and \eqref{eq:P3_full}. 
    Dot-dashed curves are based on Eq.~(\ref{eq:PkAsymptotic}) extrapolated to $E=0$.
    (b) Exact cumulative probabilities $P_{\rm R}(E,t)$ of $^{87}$Rb sensor atoms as functions of $E$ at several hold times $t$ based on Eq.~\eqref{eq:prob} using the exact expressions for $P_k(T_{\rm S},E_{\rm c},E)$.
    From top to bottom colors encode  $Knt=0$ to $2.5$ in steps of $0.25$.
    System parameters are the same as those in panel (a). 
    (c) Fractional remaining sensor-atom number $\eta_{\rm S}(t)=P_{\rm R}(W,t)$ as functions of $Knt$ with trap depth $W=1.5940$~mK and other system parameters as in panel (a).
    The solid black curve corresponds to Eq.~\eqref{eq:prob} with the exact expressions for $P_k(T_{\rm S},E_{\rm c},E)$.
    The approximation in Eq.~(\ref{eq:BesselI}) is shown as the dashed purple curve.
    The solid orange curve shows the approximate prediction of Eq.~\eqref{eq:exponential}. 
    Finally, the solid blue curve shows $\eta_{\rm S}(t)=e^{-Knt}$ for the ``ideal'' CAVS, where $W\rightarrow 0$.
     }
    \label{fig:probabilities}
\end{figure*}

\section{Visualization of the probabilistic model}

Figure~\ref{fig:probabilities}(a) shows the exact and approximate cumulative probabilities $P_k(T_{\rm S},E_{\rm c},E)$ scaled by $(\alpha_1 E)^k$ for $k=1$, 2 and 3 as functions of $E$ for $^{87}$Rb sensor atoms and a Xe background gas  using  $\alpha_1$ and $\alpha_2$ derived from data in Ref.~\onlinecite{Klos2023}, reproduced in our Table \ref{tab:Rb_results}, with parameters $\delta=3/2$, $E_{\rm c}/k_{\rm B}=0.2$~mK and $T_{\rm S}=53$~$\mu$K that are motivated by the experiments of Ref.~\onlinecite{Barker2023} and discussed in App.~\ref{sec:app:temperature}.
For the largest $E$ shown in the figure, $P_A(E) \approx \alpha_1E \approx 0.2$.
We find that, for $E\gg E_{\rm c}$ and all three $k$, the approximate expression in Eq.~(\ref{eq:PkAsymptotic}) approaches the exact expression as expected.
Moreover, $P_k(T_{\rm S},E_{\rm c},E/[P_A(T_{\rm S},E_{\rm c},E)]^k<1$ for all $k$ and $E$ shown, representing a breakdown of the approximation $P_k(T_{\rm S},E_{\rm c},E)=[P_{\rm A}(E)]^k$ and, consequently, assumptions (1) and (2) described in the introduction.

Figure~\ref{fig:probabilities}(b) shows $P_{\rm R}(E,t)$, computed from  Eq.~\eqref{eq:prob} and our exact expressions for $P_k(T_{\rm S},E_{\rm c},E)$, as functions of $E$ for various hold times $t$.
As in Ref.~\onlinecite{Stewart2022}, the curves have a non-continuous derivative at $E=E_{\rm c}$.
For $E<E_{\rm c}$, $P_{\rm R}(t,E)$ is mostly determined by the initial cumulative distribution $P_0(T_{\rm S},E_{\rm c},E)$ of atoms in the trap and the probability with time that these atoms have not undergone a collision $p_0^{(c)}(t)$.
For $E>E_{\rm c}$, $P_{\rm R}(t,E)$ is independent of $E$ at $t=0$, indicating no population at these total energies, while for $t>0$, it acquires an increasing positive slope ${\rm d}P_{\rm R}(t,E)/{\rm d}E$ with $t$, indicating the build up of hotter atoms that have experienced one or more glancing collisions.

The exact remaining fraction of atoms $\eta_{\rm S}(W,t)=P_{\rm R}( W,t)$ as well as several approximations are shown in Fig.~\ref{fig:probabilities}(c) with $W=1.5940(2)$~mK and other system parameters as in panels (a) and (b).
The exact curve for $\eta_{\rm S}(W,t)$ lies significantly above the  exponential loss curve
$e^{-Knt}$ of the  ``ideal'' CAVS in the limit $W\rightarrow 0$
and lies only slightly below the exponential loss curve of Eq.~(\ref{eq:exponential}).
The non-exponential prediction of Eq.~(\ref{eq:BesselI}) is nearly indistinguishable from the exact result \eqref{eq:prob}, with about a 1.5~\% error at $Knt=6$.

To further validate our analytical calculations, we perform Monte-Carlo simulations of sensor atom loss.
The simulations begin at $t=0$ with $10^6$ sensor atoms at initial energies chosen at random according to $P_0(T_{\rm S},E_{\rm c},E)$ in Eq.~(\ref{eq:PI}).
After every time step $\Delta t=0.005 \times (1/K n)$, chosen such that $ K n \Delta t\ll1$, a random number in the interval [0,1) is chosen for each atom and if that number is smaller than $ K n \Delta t$, then the sensor atom undergoes a collision with a background gas molecule.
For  sensor atoms undergoing collisions, a random energy $\Delta E$ according to the probability density in Eq.~(\ref{eq:p_I}) is chosen and is added to the sensor atom's current energy.
If a sensor atom's new energy is larger than $W$, the sensor atom is ejected, {\it i.e.} removed from the ensemble of sensor atoms in the Monte-Carlo simulations.
Cumulative probability distributions $P^{({\rm MC})}_k(T_{\rm S},E_{\rm c},E)$, $P^{({\rm MC})}_{\rm R}(E,t)$, and $\eta^{({\rm MC})}_{\rm S}(W,t)$ 
for the Monte-Carlo simulations are then constructed.
The Monte Carlo simulations and our exact results agree; {\it i.e.,} the curves for the system 
parameters used in Fig.~\ref{fig:probabilities} can not be distinguished on the scales of the three panels
in the figure.
 
\section{Effect on the recent comparison between classical and quantum standards}
\label{sec:effect}

Decay curves $\eta_{\rm S}(t)$ in the experiments of Ref.~\onlinecite{Barker2023} were measured for both $^7$Li and $^{87}$Rb sensor atoms colliding with near-room-temperature $X={\rm He}$, Ne, N$_2$, Ar, Kr, and Xe background gases introduced into the CAVSs at a known number density $n$.
The time evolution of the decay curves was then fit to an exponential with loss rate $\Gamma$, which was assumed to be of the form $\Gamma = L n = Kn[1 - P_{\rm A}(W)]$ with $P_{\rm A}(W) = (a_{\rm gl} W - b_{\rm gl} W^2)/K$.
This form for $\Gamma$ is only valid  under assumptions (1) and (2), described in the introduction.

Let us first consider the effect of violating these assumptions on the $^7$Li-$X$ collision data, which were taken at  trap depth $W=k_{\rm B} \times 0.95(14)$~mK, leading to $\alpha_1W\ll 1$ as the $\alpha_1$ for $^7$Li are two orders of magnitude smaller than those for $^{87}$Rb sensor atoms.
The measured exponential decay rates $\Gamma$ as functions of $n$ were fit assuming $\Gamma = L n$ to extract an observed rate coefficient $L$. The fitted $L$ was then compared to the predicted value obtained with  quantum mechanical scattering calculations of Ref.~\onlinecite{Klos2023}.
Both theory and experiment relied on assumptions (1) and (2).
We thus need to correct the theoretical predictions of $L$ given the finite $T_{\rm S}$ and characterize potential systematic shifts due to non-exponential decay on the extracted $L$ for $^7$Li.

To update the theoretical predictions of $L$, we need an accurate estimate of the temperature $T_{\rm S}$ for $^7$Li.
A convenient value of $T_{\rm S}=0.1$~mK was used in Ref.~\onlinecite{Barker2023}, which was not necessarily reflective of the actual temperature. In fact, $T_{\rm S}$ is not well known. Here, we take a much larger $^7$Li temperature of $T_{\rm S}=0.75$~mK from a temperature measurement of such atoms in a MOT in a similar apparatus.\cite{Barker2019}
We conservatively assume a standard uncertainty of $0.25$~mK and observe that $k_{\rm B}T_{\rm S}\approx W$.
We also take $E_{\rm c}=W$, and, since the last time the cloud was in thermal equilibrium was in the MOT, we 
have $\delta=3/2$.

\begin{table}
\begin{tabular}{lD{.}{.}{1.5}D{.}{.}{1.5}D{.}{.}{1.2}}
\hline\hline
System  & \multicolumn{1}{c}{$L$ (thr)} & \multicolumn{1}{c}{$L$ (exp)} & \multicolumn{1}{c}{$E_n(L)$} \\
 & \multicolumn{1}{c}{($10^{-9}$cm$^3$/s)} &  \multicolumn{1}{c}{($10^{-9}$cm$^3$/s)} \\
\hline
 $^7$Li-$^4$He & 1.66(4) & 1.72(3) & 0.63 \\
 $^7$Li-Ne & 1.55(14) & 1.634(16) & 0.28 \\
 $^7$Li-N$_2$ & 2.65(2) & 2.67(3) & 0.33 \\
 $^7$Li-Ar & 2.34(1) & 2.38(2) & 0.95 \\
 $^7$Li-Kr & 2.150(7) & 2.20(3) & 0.65 \\
 $^7$Li-Xe & 2.25(2) & 2.22(3) & -0.45 \\
\hline
\end{tabular}
\caption{Theoretical (thr) and updated experimentally (exp) determined loss rate coefficients $L$ for various natural abundance gases colliding with ultracold $^{7}$Li sensor atoms. 
The theoretical values are based on the exponential time evolution in Eq.~(\ref{eq:exponential}) with $T_{\rm S}=0.75(25)$ mK, $E_{\rm c}=W=k_{\rm B}\times 0.95(14)$ mK, and $K$, $a_{\rm gl}$ and $b_{\rm gl}$ from Ref.~\onlinecite{Klos2023}.
The last column shows the degree of equivalence $E_n(L) = (L_{\rm exp} - L_{\rm thr})/[2 u(L_{\rm exp}-L_{\rm thr})]$. 
All uncertainties are one-standard deviation $k=1$ uncertainties.}
\label{tab:Li_result}
\end{table}

Next, we realize that for the special case $E_{\rm c}=W=k_{\rm B}T_{\rm S}$ and $\delta=3/2$ the cumulative probabilities
$P_k(T_{\rm S},E_{\rm c},W)$ are
\begin{eqnarray}
    P_1(W/k_{\rm B},W,W)&=& \frac{11-4e}{2e-5} \alpha_1 W + O(W^2)
   \label{eq:P1WW} \\
    &=& 0.290\cdots \alpha_1 W  + O(W^2)
    \nonumber
\end{eqnarray}
and 
\begin{eqnarray}
P_2(W/k_{\rm B},W,W) &=&  \frac{7e-19}{2e-5} (\alpha_1 W)^2 +O(W^3)
 \\
 &=&0.0640\cdots (\alpha_1 W)^2  +O(W^3) \,.
 \nonumber
\end{eqnarray} 
Equation (\ref{eq:P1WW}) implies that with $k_{\rm B}T_{\rm S}\approx E_{\rm c}= W$, as in our $^7$Li experiments, $P_1(T_{\rm S},W,W)$ is about three times smaller than $P_{\rm A}(W)$. This can be compared 
 to $P_1(T_{\rm S},E_{\rm c},W)=P_{\rm A}(W)$ when $k_{\rm B}T_{\rm S}\ll E_{\rm c}<W$ analyzed in the previous section.
Thus, glancing collisions are  less of a concern than assumed in Ref.~\onlinecite{Barker2023}.
Physically, because $k_{\rm B}T_{\rm S}\approx W$, it is easier for $^7$Li atoms to be ejected from the trap.

Secondly, our measure for non-exponential behavior  $\Xi(W,t)=| P_2-P_1^2|(Knt)\approx 0.02 (\alpha_1W)^2Knt$  is at most $10^{-5}$ for all background gases at the largest $Knt=4$ measured in Ref.~\onlinecite{Barker2023}. 
Consequently, at our level of accuracy for extracting loss rates $\Gamma$ for $^7$Li, contributions from non-exponential decay can be neglected and the rate coefficient 
$L= K[1-P_1(T_{\rm S},E_{\rm c},W)]$ with $ P_1(T_{\rm S},E_{\rm c},W)$ from Eq.~(\ref{eq:P1WW}) can be used. 
Because $0<P_1(T_{\rm S},E_{\rm c},W)<P_{\rm A}(W)$, the theoretically predicted values for $L$ are larger than those in Ref.~\onlinecite{Barker2023}.
Moreover, the contribution to the uncertainty of $L$ from $W$ is reduced compared to that in Ref.~\onlinecite{Barker2023}.
The updated predictions for $L$, shown in Table~\ref{tab:Li_result}, are now in better agreement with their experimental counterparts, with the theoretical and experimental values agreeing at two standard deviations ($k=2$).

An updated uncertainty budget for $^7$Li-Ar and the change in the theoretical predictions of $L$ from Ref.~\onlinecite{Barker2023} are shown in the Supplemental Tables~S1 and~S2, respectively.
The maximum change in $L$ from Ref.~\onlinecite{Barker2023} is 0.64~\% for $^7$Li+Xe. In fact, for all background gases the change in $L$ is less than our updated corresponding theoretical uncertainty.

The situation is more complicated for $^{87}$Rb sensor atoms.
Once again, we require a better estimate of the sensor-atom temperature $T_{\rm S}$. It is now obtained by measuring the energy distribution of the $^{87}$Rb  atoms in the quadrupole trap as in Ref.~\onlinecite{Stewart2022} and  described in App.~\ref{sec:app:temperature}.
We find $T_{\rm S}=53(12)$~$\mu$K ensuring that we are in the limit $k_{\rm B}T\ll E_{\rm c}<W$ studied in Sec.~\ref{sec:analysis}.

In Ref.~\onlinecite{Barker2023}, exponential decay was assumed for the loss of $^{87}$Rb sensor-atoms. For $k_{\rm B}T\ll E_{\rm c}<W$,
the measure of non-exponential behavior  $\Xi(W,t)\approx 0.50  (\alpha_1W)^2 Knt$
is largest for a Xe background gas at $\Xi(W,t)=0.028$ using the longest experimental hold time. This value is comparable to the fractional standard statistical uncertainty $u(\Gamma)/\Gamma$ for the decay rates $\Gamma$ found from exponential fits to $\eta_{\rm S}(t)$ versus $t$.
Thus, we refit the data of Ref.~\onlinecite{Barker2023} to  account for  non-exponential decay.

For our fit, let us first consider the exact solution in Eq.~$\eqref{eq:prob}$ with the substitutions of $\eta_{\rm S}(t) = \eta_0 P_{\rm R}(W,t)$ and $\Gamma_0 = Kn$, which yields
\begin{equation}
    \label{eq:prob2}
    \eta_{\rm S}(t) = \eta_0 e^{-\Gamma_0 t}\left[1 + P_1(\Gamma_0 t) + \frac{1}{2} P_2 (\Gamma_0 t)^2 + \frac{1}{6} P_3(\Gamma_0 t)^3 + \cdots\right]\,. 
\end{equation}
with probabilities $P_k(\cdot,\cdot,\cdot)$ evaluated at $E=W$.
One immediate question arises: how many $P_k(T_{\rm S},E_{\rm c},W)$ must be included given our measurement uncertainties in $\eta_{\rm S}(t)$?
Using a second convenient substitution of $\Gamma_i = K n \alpha_i = \Gamma_0 \alpha_i$ for $i=1$, 2, $\dots$ together with our approximate expression for $P_k(T_{\rm S}, E_{\rm c}, W)$
in Eq.~\eqref{eq:PkAsymptotic}, we find
\begin{eqnarray}
  \lefteqn{  \eta_{\rm S}(t)  \approx  e^{-\Gamma_0 t}\left\{1 + \sum_{i=1}^2 \Gamma_i W^i \left(1-i (3/2+\delta)\frac{k T_{\rm S}}{W}\right) t  \right.}  \label{eq:approx_Rb_fit_function_2} \\ 
    & & \left. + \left[\frac{1}{4} (\Gamma_1 W)^2 + \frac{1}{3}\Gamma_1\Gamma_2 W^2 \right] t^2  + \frac{1}{36} (\Gamma_1 W)^3 t^3  + \cdots\right\} \,. 
    \nonumber
\end{eqnarray}
Thus, we observe that the dominant coefficient for each $t^k$ in the braces $\{\cdots\}$ is $(\Gamma_1 W)^k/[\Gamma(k+1)]^2$.
For the data of Ref.~\onlinecite{Barker2023}, the maximum $W$ for which data was taken is $W/k_{\rm B}=1.6$~mK and the maximum $t$ was approximately $ 3/\Gamma_0$.
At $t=3/\Gamma_0$, the measured relative uncertainty $u(\eta_{\rm S}(t))/\eta_{\rm S}(t)\gtrsim \epsilon\equiv 0.05$.
We then demand that $ (3\Gamma_1W/\Gamma_0)^k/[\Gamma(k+1)]^2<0.01\epsilon$ to ensure convergence of the series in Eq.~\eqref{eq:prob2}.
Using the theoretical prediction for $^{87}$Rb-Xe of Ref.~\onlinecite{Klos2023}, which has the largest $\Gamma_1$ of all studied background gases, this requirement demands computing terms up to and including $k=4$.

The data of Ref.~\onlinecite{Barker2023} is further complicated by apparent two-body losses.
We thus fit the normalized $^{87}$Rb atom number $\eta_{\rm S}(W,t)$ as a function of $t$ and $W$ for given $T_{\rm S}$, $E_{\rm c}$, and background gas number density $n$ to the numerical solution of 
\begin{equation}
    \label{eq:loss_2}
    \frac{{\rm d}\eta_{\rm S}}{{\rm d}t} =
    -\Gamma(W, t) \eta_{\rm S} - \beta(W) \eta_{\rm S}^2\ ,
\end{equation}
with $\eta_{\rm S}(t=0) = \eta_0$, where the trap depth and time dependent $\Gamma(W,t)$ is 
\begin{eqnarray}
    \Gamma(W, t) & = & -\frac{{\rm d} \log P_{\rm R}(W,t)}{{\rm d}t} =  -\frac{1 + \sum_{i=1}^4  P_k t^{k-1}/\Gamma(k)}
     {1 + \sum_{i=1}^4 
     P_k t^k/\Gamma(k+1)}
     \,, \label{eq:log_derivative_2}
\end{eqnarray}
and the two-body loss parameter $\beta(W)=\beta_0+\beta_1 W$.
We have verified that the numerical solution of Eq.~\eqref{eq:loss_2} using the second equality of Eq.~\eqref{eq:log_derivative_2} reproduces Eq.~\eqref{eq:prob2} to better than $0.0001 \epsilon$, where $\epsilon=0.05$ as before.

For each value of $n$, the authors of Ref.~\onlinecite{Barker2023} measured time traces $\eta_{\rm S}(W,t)$ between $t=0$ and $t\approx 4/\Gamma_0$ for seven $W$ between $k_{\rm B}\times0.4$~mK and $k_{\rm B}\times1.6$~mK.
The sensor atom temperature and  cutoff energy are fixed at $T_{\rm S}=53(12)$~$\mu$K and $E_{\rm c}=k_{\rm B}\times 0.1993(2)$ mK, respectively.
Moreover, we assume $\delta=3/2$.
The quality of the experimental data determines the number of $\Gamma_i$ included in the fit. We use $\Gamma_i$ for $i=1,\cdots,i_{\rm max}$, where $i_{\rm max}$ is the first value of $i$ that satisfies $\Gamma_{i_{\rm max}}<2 u(\Gamma_{i_{\rm max}})$ and $u(\Gamma_i)$ is the statistical standard uncertainty in $\Gamma_i$.
Consequently, for $^{87}$Rb-He the adjusted parameters are $\eta_0$, $\Gamma_0$, $\Gamma_1$, $\beta_0$, and $\beta_1$.
For all others systems, $\Gamma_2$ is the sixth adjusted parameter. 
We propagated the 20\,\% uncertainty of $T_{\rm S}$ through the fits by determining ${\rm d} \Gamma_i/{\rm d} T_{\rm S}$ for $i=0,\cdots,i_{\rm max}$.
The $<0.1\,\%$ uncertainties of $E_{\rm c}$ and $W$ are negligible contributions to our total uncertainty.
The fitted $\beta(W)>0$ at a 2-$\sigma$ ($k=2$) uncertainty for 33~\% of our time traces, possibly indicating the presence of two body collisional induced loss. For the remaining fits, $\beta(W)$ is consistent with zero at 2-$\sigma$.

By construction for each background gas, fitted quantities $\Gamma_i$ with $i=0, \cdots, i_{\rm max}$ should only be proportional to the background gas number density $n$.
We, however, observe offsets as function of $n$ and thus obtain rate coefficients $K$ and glancing collision rates $a_{\rm gl}$ and $b_{\rm gl}$ for the roughly 10 values of $n$ between $4\times10^{7}$~cm$^{-3}$ and $4\times10^{9}$~cm$^{-3}$ using
\begin{eqnarray}
    \Gamma_0 & = & Kn \,+ \Gamma_{0,\rm base} \nonumber\,, \\
    \Gamma_1 & = & a_{\rm gl} n \,+ \Gamma_{1,\rm base}\,, \\
    \Gamma_2 & = & b_{\rm gl} n \,+ \Gamma_{2,\rm base} \nonumber\,.
\end{eqnarray}
with offsets $\Gamma_{i,\rm base}$ as two additional adjusted parameters for $^{87}$Rb-He and three additional adjusted parameters for all others background gases.
In these second, linear fits, we have accounted for the  uncertainties of and covariances among the two or three $\Gamma_i$ obtained by fitting the time traces as well as the roughly 0.3~\% relative uncertainty in $n$.
Rate $\Gamma_{0,\rm base}$ is $0.027(6)$~s$^{-1}$ independent of the background gas species 
$\Gamma_{1,\rm base}$ and $\Gamma_{2,\rm base}$ are consistent with zero at 2-$\sigma$ in 70~\% of the fits and at 3-$\sigma$ for all of the fits.
It is likely that $\Gamma_{0,\rm base}$ is caused by residual gas, likely H$_2$, at our base or lowest pressure.
Assuming H$_2$ gas with $K_{{\rm H}_2}=3.9(1)\times10^{-9}$~cm$^3$/s\cite{Klos2023}, we derive $n_{\rm base}=\Gamma_{0,\rm base}/K_{{\rm H}_2} = 6.9(1.5)\times10^6$~cm$^3$/s, corresponding to base pressure $p_{\rm base} = n_{\rm base} k_{\rm B}T = 2.8(6)\times10^{-8}$~Pa, in agreement with those extracted from decay curves measured at $n=0$ in Ref.~\onlinecite{Barker2023}.
If the $\Gamma_{i,\rm base}$ for $i=1$ and 2 are also caused by H$_2$ gas and if $\Xi(W,t)\ll 1$ for all $t$ for both the residual gas and the species of interest, then we can show that $\Gamma_{1,\rm base}=a_{\rm gl, base} n_{\rm base}$ and $\Gamma_{2,\rm base} = b_{\rm gl, base} n_{\rm base}$.
The corresponding fit values of $a_{\rm gl}$ and $b_{\rm gl}$ are consistent with the expected values of $a_{\rm gl}\approx 1.4\times10^{-7}$~cm$^3$/(s~K) and $b_{\rm gl}\approx 8\times10^{-5}$~cm$^3$/(s~K$^2$) for H$_2$, but also consistent with zero within at a 2-$\sigma$ ($k=2$) uncertainty.

Table~\ref{tab:Rb_results} shows the comparison between the updated $^{87}$Rb experimental values for $K$, $a_{\rm gl}$, and $b_{\rm gl}$ and the corresponding original theoretical values of Ref.~\onlinecite{Klos2023}. Generally the experimental values are now in better agreement with the theory.
As with the results of Ref.~\onlinecite{Barker2023}, all numbers agree at two-standard deviations ($k=2$) except $K$ for $^{87}$Rb-Ar, which agrees only at four-standard deviations ($k=4$), and $a_{\rm gl}$ for $^{87}$Rb-Kr, which agrees at three standard deviations ($k=3$).
Supplemental Tables S3, S4, and S5 show a sample uncertainty budget for $^{87}$Rb-Ar, the change in the values of $K$, $a_{\rm gl}$, and  $b_{\rm gl}$ from Ref.~\onlinecite{Barker2023} to this work, and the current state of knowledge for values of $K$, respectively.
The maximum relative change in $K$ is $-2.2$~\% for $^{87}$Rb-Xe, as expected given that Xe results in the most glancing collisions.

\begin{table*}
\begin{tabular}{lD{.}{.}{1.4}D{.}{.}{1.4}D{.}{.}{3.4}D{.}{.}{1.5}D{.}{.}{1.6}D{.}{.}{3.4}D{.}{.}{1.5}D{.}{.}{1.4}D{.}{.}{3.4}}
\hline\hline
System  & \multicolumn{1}{c}{$K$ (thr)} & \multicolumn{1}{c}{$K$ (exp)} & \multicolumn{1}{c}{$E_n(K)$} & \multicolumn{1}{c}{$a_{\rm gl}$ (thr)} & \multicolumn{1}{c}{$a_{\rm gl}$ (exp)} & \multicolumn{1}{c}{$E_n(a_{\rm gl})$} & \multicolumn{1}{c}{$b_{\rm gl}$ (thr)} & \multicolumn{1}{c}{$b_{\rm gl}$ (exp)} & \multicolumn{1}{c}{$E_n(b_{\rm gl})$} \\
 & \multicolumn{1}{c}{($10^{-9}$cm$^3$/s)} &  \multicolumn{1}{c}{($10^{-9}$cm$^3$/s)} & & \multicolumn{1}{c}{($10^{-7}$cm$^3$/[s K])} & \multicolumn{1}{c}{($10^{-7}$cm$^3$/[s K])} & & \multicolumn{1}{c}{($10^{-5}$cm$^3$/[s K$^2$])} & \multicolumn{1}{c}{($10^{-5}$cm$^3$/[s K$^2$])} & \\
\hline
 $^{87}$Rb-$^4$He & 2.37(3) & 2.35(6) & -0.18 & 0.336(5) & -0.42(85) & 0.44 & 0.067(3) & \multicolumn{1}{c}{---} & \multicolumn{1}{c}{---} \\
 $^{87}$Rb-Ne & 2.0(2) & 2.21(5) & 0.54 & 1.06(9) & 1.2(7) & -0.09 & 0.59(3) & 2.0(3.5) & -0.21 \\
 $^{87}$Rb-N$_2$ & 3.45(6) & 3.56(8) & 0.54 & 2.6(2) & 1.9(1.4) & 0.24 & 2.3577(7) & 5.8(7.2) & -0.24 \\
 $^{87}$Rb-Ar & 3.035(7) & 3.29(5) & 2.38 & 2.42(2) & 2.6(8) & -0.11 & 2.19(2) & -2.4(3.9) & 0.60 \\
 $^{87}$Rb-Kr & 2.787(10) & 2.80(4) & 0.23 & 3.04(2) & 1.8(5) & 1.18 & 3.97(3) & 3.8(2.5) & 0.02 \\
 $^{87}$Rb-Xe & 2.880(10) & 2.87(6) & -0.08 & 4.11(5) & 3.5(1.0) & 0.30 & 7.1(1) & 12.8(4.5) & -0.64 \\

\hline
\end{tabular}
\caption{
Theoretical~\cite{Klos2023}  (thr) and updated experimental (exp)  values for the loss rate coefficient $K$ at zero trap depth, the first-order glancing rate coefficient $a_{\rm gl}$, and the second-order glancing rate coefficient $b_{\rm gl}$ for various natural abundance  gases colliding with ultracold $^{87}$Rb atoms.
Numbers in parentheses are one-standard-deviation, $k=1$ uncertainties.
The degree of equivalence is $E_n(K) = (K_{\rm exp} - K_{\rm thr})/[2 u(K_{\rm exp}-K_{\rm thr})]$ for $K$ and likewise for $a_{\rm gl}$ and  $b_{\rm gl}$.
}
\label{tab:Rb_results}
\end{table*}

\section{Conclusion}

We have developed a probabilistic and analytic model of glancing collisions in a CAVS.
While inspired by the recent semiclassical results of Ref.~\onlinecite{Shen2021, Stewart2022}, our model can use either semiclassical scattering theory\cite{Booth2019, Shen2020, Shen2021, Stewart2022, Shen2023} or fully quantum mechanical scattering theory\cite{Makrides2019,Makrides2020, Makrides2022Ea, Makrides2022Eb, Klos2023} as an input.
We use our model to update the values of Ref.~\onlinecite{Barker2023} based on a more complete description of the time evolution of the number of ultracold sensor atoms.
We find that our relative adjustments for the theoretically predicted $L$ for $^7$Li-X are $\lesssim 0.6$~\%.
For $^{87}$Rb-X, we find that our relative adjustments are as large as $2.2$~\% for our experimental, extrapolated zero-trap depth loss rate coefficients, $K$.
Importantly, we note that a CAVS based on $^7$Li is far less sensitive to the initial temperature of the cold atom cloud and the effects of glancing collisions than one based on $^{87}$Rb.
A CAVS based on $^{87}$Rb can compensate for such complications, but requires ancillary measurements to achieve high levels of accuracy.

\section*{}
\appendix

\section{Collision kinematics and timescales}
\label{sec:app:kinematics}

In this appendix, we briefly review the kinematics of collisions in the context of ultracold sensor atoms of mass $m$ held in a trap with potential energy $U(\mathbf{r})$ at location $\mathbf{r}$ colliding with room-temperature background atoms or molecules.
The minimum potential energy occurs at $\mathbf{r}=\mathbf{0}$ and $U(\mathbf{r}=\mathbf{0})=0$.
Therefore, total energy $E=0$ is the lowest energy of an atom in the trap, which occurs when it is at rest at $\mathbf{r}=\mathbf{0}$.
Atoms with $E>0$ execute classical orbits starting from position $\mathbf{r}$ and velocity $\mathbf{v}$ such that $E=m|{\bf v}|^2/2+U({\bf r})$ is conserved. For large $\mathbf{r}$, potential $U(\mathbf{r})$ approaches
trap depth $W$ from below along at least one direction $\mathbf{r}$.
Typical timescales for orbits of atoms with magnetic moment $\mu$ in a quadrupole magnetic trap are  $\tau_{\rm orbit} \sim \sqrt{2 m k_{\rm B} T_{\rm S}/(\mu B')^2}$, where atom temperature $T_{\rm S}\ll W/k_{\rm B}$, $\mu\approx\mu_B$, $\mu_{\rm B}$ is the Bohr magneton, and $B'$ is the  magnetic field gradient of the quadrupole trap. In our CAVSs, $\tau_{\rm orbit}$ is of the order of 1~ms. 
If an atom is executing an orbit with $E>W$, it is not bound by the trap and is ejected.

The sensor atom and background particle do not move appreciably during the collision. This
can be seen by noting that the duration of a collision is $\tau_{\rm collision}\sim \sqrt{\sigma_{\rm eff}}/\langle v \rangle$,
where $\langle v\rangle$ is the mean velocity of the background gas atom or molecule  at temperature $T$
and $\sigma_{\rm eff}=K/\langle v\rangle\equiv\pi d_{\rm eff}^2$ is the effective cross section of the collision.
That is, $\tau_{\rm collision}$ is related to the time it takes to traverse the effective diameter $d_{\rm eff}$ of the collision
partners  near room temperature.
Given that  $K\sim 10^{-9}$~cm$^3$/s and $\langle v \rangle\sim 10^4$~cm/s, we find that $\tau_{\rm collision}\sim 10\ {\rm ps} \ll \tau_{\rm orbit}$.

A consequence of the large difference in timescales $\tau_{\rm collision}$ and $\tau_{\rm orbit}$ is that collisions occur at well-defined locations.
Thus, the change in energy for a sensor atom with initial velocity $\mathbf{v}_{\rm S}$  and final velocity $\mathbf{v}_{\rm S}+\delta \mathbf{v}_{\rm S}$ colliding at location $\mathbf{r}$ with a background gas particle is   
\begin{eqnarray}
    \Delta E & =&  U(\mathbf{r}) + \frac{1}{2}m (\mathbf{v}_{\rm S}+\delta \mathbf{v}_{\rm S})^2 - \left[U(\mathbf{r}) + \frac{1}{2}m \mathbf{v}_{\rm S}^2\right] \nonumber \\
    & = & \frac{1}{2} m (\delta \mathbf{v}_{\rm S})^2 + m \mathbf{v}_{\rm S}\cdot\mathbf{\delta v}_{\rm S}\,. \label{eq:kinematics} 
\end{eqnarray}
For most collisions and even most glancing collisions $|\delta {\bf v}_{\rm S}| \gg |{\bf v}_{\rm S}|$ as 
$ T_{\rm S}\ll T$ and $k_{\rm B} T_{\rm S}\ll W$, respectively.
Furthermore, the bombardment from background gas particles is isotropic and thus the second term in Eq.~(\ref{eq:kinematics}) is zero when averaged over all possible orientations of $\delta{\bf v}_{\rm S}$.
We conclude that the  ``average'' collision adds positive energy $\Delta E = m (\delta \mathbf{v}_{\rm S})^2/2$ to the sensor atom's orbit
and that most glancing collisions with background gas particles, {\it i. e.} those with $\Delta E<W$, heat the sensor-atom cloud.

\section{Refined estimate of the temperature of ultracold $^{87}$Rb samples}
\label{sec:app:temperature}

\begin{figure}
    \centering
    \includegraphics{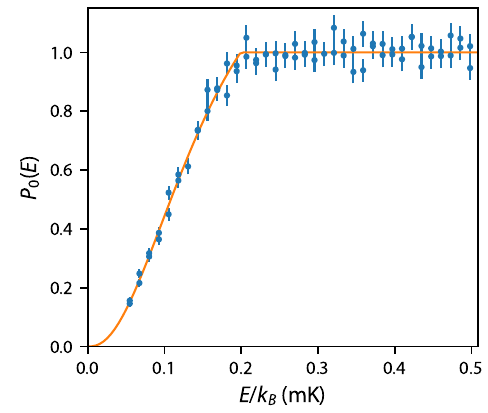}
    \caption{Experimentally measured $P_0(T_{\rm S},E_{\rm c},E)$ with standard uncertainties (blue filled circles) for $^{87}$Rb sensor atoms as a function of total energy $E$
    with cutoff energy $E_{\rm c}=k_{\rm B}\times 0.1993(2)$ mK.  
    The solid orange curve is a fit of the data to Eq.~(\ref{eq:PI}) with temperature 
    $T_{\rm S}=47(7)$~$\mu$K at a fixed value $\delta=3/2$.
    }
    \label{fig:initial_distribution}
\end{figure}

In this appendix, we describe an updated temperature measurement $T_{\rm S}$ of the ultracold $^{87}$Rb cloud.
We use a technique similar to that in Ref.~\onlinecite{Stewart2022}.
Loss rate coefficients between various gases and $^{87}$Rb were measured in the laboratory-scale cold-atom vacuum sensor (l-CAVS).
As described in Refs.~\onlinecite{Barker2022, Barker2023}, the l-CAVS contains an ``RF knife,'', which applies radio-frequency magnetic fields to the atoms to induce spin-flip transitions at specific locations in a magnetic trap defined by the RF frequency $\nu_{\rm RF}$ and local Zeeman shift in the trap.
In Ref.~\onlinecite{Barker2023}, the RF knife was used to set cutoff energy $E_{\rm c}$ and trap depth $W$.
To set $E_{\rm c}$, a frequency ramp from $\nu_{\rm RF}^{\rm init}=40$~MHz to $\nu^{\rm final}_{\rm RF}=5$~MHz in 1 s removed atoms from the trap with total energy $E>E_{\rm c}=h \nu^{\rm final}_{\rm RF} \{1-2mg/(\mu_{\rm B} B')\} = k_{\rm B} \times 0.1993(2)$~mK, where $h$ is Planck's constant, $m$ is the mass of $^{87}$Rb, $g$ is the predicted local gravitational acceleration from Ref.~\onlinecite{NOAASurfaceGravityPrediction}, and $B'$ is the measured axial magnetic field gradient of the quadrupole trap.
The uncertainty in $E_{\rm c}$ is limited by our knowledge of $B'$.
Immediately after the ramp, the RF frequency was quickly (essentially instantaneously) changed to a constant, larger value to set trap depth $W$ for the remainder of the measurement.

In this work, we extend our use of the RF knife to measure the initial total-energy distribution $P_0(T_{\rm S},E_{\rm c},E)$
 of $^{87}$Rb atoms in the quadrupole magnetic trap as function of total energy $E$.
We follow the procedure  described above to prepare the cloud with $E_c= k_{\rm B} \times 0.1993(2)$~mK and subsequently increase the RF frequency to 10~MHz leading to the smallest $W$ used in 
Ref.~\onlinecite{Barker2023}. In fact, $W=k_{\rm B}\times 0.3986(4)$~mK.
After holding the atoms in the trap for 10~s at this $W$, a hold time short compared to typical vacuum lifetimes of roughly $150$~s and the typical Rb-Rb collisional thermalization time of roughly 30~s, frequency $\nu_{\rm RF}$ is increased to 40~MHz and subsequently slowly decreased to a value between 100~kHz and 12.5~MHz, ejecting all $^{87}$Rb atoms with energy $E> h\nu_{\rm RF}(1-2mg/\mu_B B')$.
Finally, we recapture the remaining sensor atoms  into a MOT to count them to give $N(E)$ as in Ref.~\onlinecite{Barker2023}.
In this way, we measure the distribution $P_0(T_{\rm S},E_{\rm c},E)=N(E)/N_{\rm init}$ between $E = k_{\rm B}\times 3.986(4)$~$\mu$K and $k_{\rm B}\times 0.4982(4)~{\rm mK} \equiv E_{\rm max}$, where $N_{\rm init}$ is the sensor atom number just before the frequency of the RF knife is slowly decreased from 40~MHz, which corresponds to $E=E_{\rm init}\equiv k_{\rm B} \times 1.5940(2)$~mK.
Standard uncertainties in sensor atom number $N(E)$ are taken as in Ref.~\onlinecite{Barker2023}, with the $u(N)=\sqrt{(\sigma_N N^2)+\sigma_0^2}$, with $\sigma_N=0.04$ and $\sigma_0 = 300$, yielding reduced $\chi^2_\nu$ values near unity and residuals that are independent of $E$.
In practice, as $E_{\rm max}< E_{\rm init} $ atom number $N_{\rm init}$ is taken as an adjusted constant.

The results of the measurement of $N(E)$ and thus $P_0(T_{\rm S},E_{\rm c},E)$ as a function of $E$ for  $E_{\rm c}=k_{\rm B} \times 0.1993(2)$~mK are shown in Fig.~\ref{fig:initial_distribution}.
For $E>E_{\rm c}$, $P_0(T_{\rm S},E_{\rm c},E)$ is constant, showing both the effectiveness of the initial sweep of the RF knife to remove ``hot'' sensor atoms and the lack of a significant number of Rb-Rb rethermalizing collisions during the 10~s hold time.
For the present experiment, the RF antenna is less effective at removing $^{87}$Rb atoms with $E/k_{\rm B}\lesssim 0.05$~mK, so no data are shown in that region.
A fit with $T_{\rm S}$, $\delta$, and $N_{\rm init}$ as adjusted constants yields $T_{\rm S}=59(8)$~$\mu$K and $\delta=1.13(19)$, which includes at 2-$\sigma$ our expected value of $\delta=3/2$.
A second fit, indistinguishable in Fig.~\ref{fig:initial_distribution}, with a fixed and physically motivated value of $\delta=3/2$, yields $T_{\rm S}=47(7)$~$\mu$K.
In an attempt to account for systematic effects, we take as a conservative estimate for $T_{\rm S}$ the weighted mean of the two estimated temperatures 
assuming a standard uncertainty given by the temperature difference, which gives $T_{\rm S}=53(12)$~$\mu$K.

\section*{Acknowledgements}
The authors thank K. Madison, J. Booth, and A. Deshmukh for useful discussions; K. Madison and R. Krems for organizing the workshop that helped to elucidate the physics, and P.J. Egan and B. Reschovsky for a thorough reading of the manuscript.

\section*{Author Declarations}
\subsection*{Conflicts of Interest}
D.S.B., J.A.F., J.S., and S.P.E. have U.S. patent 11,291,103 issued. D.S.B. and S.P.E. have filed U.S. provisional patent 63/338,047.

\section*{Data Availability}
The data that support the findings of this study are available from the corresponding author upon reasonable request.

\bibliography{main}

\end{document}


\title{Supplemental tables for ``On the effect of `glancing' collisions in the cold atom vacuum standard''}

\author{Stephen P. Eckel\,\orcidlink{0000-0002-8887-0320}}
\email{stephen.eckel@nist.gov}
\affiliation{\SSD}
\author{Daniel S. Barker\,\orcidlink{0000-0002-4614-5833}}
\affiliation{\SSD}
\author{James A. Fedchak\,\orcidlink{0000-0002-1259-0277}}
\affiliation{\SSD}
\author{Jacek K{\l}os\,\orcidlink{0000-0002-7407-303X}}
\affiliation{\JQI}\affiliation{\UMD}
\author{Julia Scherschligt\,\orcidlink{0000-0003-4965-0103}}
\affiliation{\SSD}
\author{Eite Tiesinga\,\orcidlink{0000-0003-0192-5585}}
\affiliation{\JQI}\affiliation{\UMD}\affiliation{\QMD}

\date{\today}

\begin{abstract}
Here, we present several additional tables of including sample uncertainty budgets, shifts in values of this work compared to past work, and updated $^{87}$Rb loss rate coefficients across multiple different methods.
\end{abstract}

\maketitle

\begin{table}
\begin{tabular}{lclr}
\begin{tabular}{lllr}
\hline\hline 
 & Type & Source & Contribution (\%)\\ 
\hline 
Experi- & B & Temperature of the CAVS, $T$ & 0.51 \\ 
\ \ mental & B & Flowmeter, $\dot{N}$ & 0.24 \\ 
 & B & Orifice area, $A$ & 0.13 \\ 
 & B & Imaging non-linearity and drifts, $\Gamma$ & 0.07 \\ 
 & B & Pressure ratio, $r_p$ & 0.05 \\ 
 & B & Orifice transmission prob., $\alpha_{\rm MC}$ & 0.02 \\ 
\hline 
 & B & Subtotal & 0.59 \\ 
 & A & Subtotal & 0.73 \\ 
\hline 
 & & Total & 0.94 \\ 
\hline\hline 
Theory & B & Temperature of the CAVS, $T$ & 0.32 \\
 & B & Theory & 0.26 \\ 
 & B & Trap Depth, $W$ & 0.02 \\ 
 & B & Temperature of sensor atoms, $T_{\rm S}$ & 0.01 \\ 
 & B & Isotopic shift & $<0.01$ \\ 
\hline 
 & & Total & 0.41 \\ 
\hline\hline 
\end{tabular}
\end{tabular}
\caption{Experimental and theoretical statistical (type-A) and systematic (type-B)   uncertainty budgets of loss rate coefficient $L$ for the p-CAVS with $^7$Li sensor atoms and a natural abundance Ar background gas.  The experimental and theoretical contributions add in quadrature to the {\it relative} uncertainty $u(L)/L$ of  $L$. The experimental and theoretical temperature contributions are correlated. See Ref.~\onlinecite{Barker2023} on how this non-zero correlation is treated.}
\label{tab:uncertainty_budget}
\end{table}

\begin{table}
\begin{tabular}{lD{.}{.}{2.7}D{.}{.}{2.7}D{.}{.}{3.2}}
\hline\hline
System  & \multicolumn{1}{c}{$L$ (Ref.~\onlinecite{Barker2023})} & \multicolumn{1}{c}{$L$ (This work)} & \multicolumn{1}{c}{\% shift}\\
 & \multicolumn{1}{c}{$10^{-9}$cm$^3$/s} &  \multicolumn{1}{c}{$10^{-9}$cm$^3$/s} \\
\hline
 $^7$Li-$^4$He & 1.661(41) & 1.662(41) & +0.06 \\
 $^7$Li-Ne & 1.55(14) & 1.55(14) & +0.23 \\
 $^7$Li-N$_2$ & 2.642(22) & 2.651(22) & +0.34 \\
 $^7$Li-Ar & 2.3359(98) & 2.3443(96) & +0.36 \\
 $^7$Li-Kr & 2.1398(71) & 2.1503(68) & +0.49 \\
 $^7$Li-Xe & 2.234(21) & 2.249(21) & +0.64 \\
\hline
\end{tabular}
\caption{
Shift in the theoretically predicted values of the loss rate coefficient $L$ for various natural abundance  gases colliding with ultracold $^{7}$Li between Ref.~\onlinecite{Barker2023} and this present work.
Numbers in parentheses are one-standard-deviation, $k=1$ uncertainties.
}
\label{tab:Li_shift}
\end{table}


\begin{table}
\begin{tabular}{lclr}
\hline\hline 
 & Type & Source & Contribution (\%) \\ 
\hline 
Experi- & B & Flowmeter, $\dot{N}$ & 0.26 \\ 
\ \ mental & B & Orifice area, $A$ & 0.13 \\ 
 & B & Imaging non-linearity and drifts, $\Gamma$ & 0.07 \\ 
 & B & Temperature of the sensor atoms, $T_{\rm S}$ & 0.06 \\ 
 & B & Pressure ratio, $r_p$ & 0.05 \\ 
 & B & Temperature of the CAVS, $T$ & 0.04 \\ 
 & B & Orifice transmission prob., $\alpha_{\rm MC}$ & 0.02 \\ 
 & B & Trap depth, $W$ & $<0.01$ \\ 
\hline 
 & B & Subtotal & 0.31 \\ 
 & A & Subtotal & 1.56 \\ 
\hline 
 & & Total & 1.59 \\ 
\hline\hline 
Theory & B & Theory & 0.23 \\ 
 & B & Temperature of the CAVS, $T$ & 0.02 \\ 
 & B & Isotopic shift & $<0.01$ \\ 
\hline 
 & & Total & 0.23 \\ 
\hline\hline 
\end{tabular}
\caption{Experimental and theoretical statistical (type-A) and systematic (type-B) uncertainty budgets of zero-trap depth loss rate coefficient $K$ for the l-CAVS with $^{87}$Rb sensor atoms and a natural abundance Ar background gas.  The experimental and theoretical contributions add in quadrature to the {\it relative} uncertainty $u(K)/K$ of $K$. The experimental and theoretical temperature contributions are correlated. See Ref.~\onlinecite{Barker2023}  on how this non-zero correlation is treated.}
\end{table}

\begin{table}
\begin{tabular}{lD{.}{.}{2.5}D{.}{.}{2.5}D{.}{.}{3.3}D{.}{.}{3.5}D{.}{.}{2.5}D{.}{.}{5.2}D{.}{.}{3.7}D{.}{.}{3.7}D{.}{.}{4.2}}
\hline\hline

System  & \multicolumn{3}{c}{$K$ ($10^{-9}$cm$^3$/s)} & \multicolumn{3}{c}{$a_{\rm gl}$ ($10^{-7}$cm$^3$/[s K])} &
\multicolumn{3}{c}{$b_{\rm gl}$ ($10^{-5}$cm$^3$/[s K$^2$])} \\

 & \multicolumn{1}{c}{Ref.~\onlinecite{Barker2023}} & \multicolumn{1}{c}{This work} & \multicolumn{1}{c}{\% shift} & \multicolumn{1}{c}{Ref.~\onlinecite{Barker2023}} & \multicolumn{1}{c}{This work} & \multicolumn{1}{c}{\% shift} & \multicolumn{1}{c}{Ref.~\onlinecite{Barker2023}} & \multicolumn{1}{c}{This work} & \multicolumn{1}{c}{\% shift} \\
 
\hline
$^{87}$Rb-$^4$He & 2.34(6) & 2.35(6) & +0.4 & 0.1(8) & -0.4(8) & -130 & \multicolumn{1}{c}{---} & \multicolumn{1}{c}{---} & \multicolumn{1}{c}{---} \\
 $^{87}$Rb-Ne & 2.23(5) & 2.21(5) & -0.7 & 1.3(7) & 1.2(7) & -6.7 & 2.6(2.4) & 2.0(3.5) & -30 \\
 $^{87}$Rb-N$_2$ & 3.60(1) & 3.56(8) & -1.0 & 2.3(1.4) & 1.9(1.5) & -21 & 13.3(8.4) & 5.8(7.2)  & -130 \\
 $^{87}$Rb-Ar & 3.30(6) & 3.29(5) & -0.6 & 2.3(8) & 2.6(8) & +7.0 & -1.0(2.9) & -2.4(3.9)  & -58 \\
 $^{87}$Rb-Kr & 2.83(4) & 2.80(4) & -0.8 & 1.9(5) & 1.8(5) & -6.3 & 3.8(2.7) & 3.8(2.5) & 0 \\
 $^{87}$Rb-Xe & 2.93(7) & 2.87(6) & -2.0 & 3.9(9) & 3.5(1.0) & -11 & 10.3(3.0) & 12.8(4.5) & +23 \\
\hline
\end{tabular}
\caption{
Shift in the experimentally determined values of the loss rate coefficient $K$ at zero trap depth, the first-order glancing rate coefficient $a_{\rm gl}$, and the second-order glancing rate coefficient $b_{\rm gl}$ for various natural abundance  gases colliding with ultracold $^{87}$Rb between Ref.~\onlinecite{Barker2023} and this present work.
Numbers in parentheses are one-standard-deviation, $k=1$ uncertainties.
}
\label{tab:Rb_results}
\end{table}


\begin{table}
\begin{tabular}{lD{.}{.}{2.6}D{.}{.}{1.4}D{.}{.}{2.6}D{.}{.}{1.4}}
\hline\hline
System  & \multicolumn{4}{c}{$K$ ($10^{-9}$cm$^3$/s)} \\
 & \multicolumn{1}{c}{UQDC\cite{Shen2021}} & \multicolumn{1}{c}{Ratiometric\cite{Shen2023}} & \multicolumn{1}{c}{Theory\cite{Klos2023}} &\multicolumn{1}{c}{This work}  \\
\hline 
 $^{87}$Rb-H$_2$ &  5.12(15) & 3.8(2) &  3.9(1) &\multicolumn{1}{c}{---} \\
 $^{87}$Rb-$^4$He &  2.41(14) & \multicolumn{1}{c}{---}  & 2.37(3) & 2.35(6) \\
 $^{87}$Rb-Ne &  \multicolumn{1}{c}{---} & \multicolumn{1}{c}{---}  & 2.0(2) & 2.21(5) \\
 $^{87}$Rb-N$_2$ & 3.14(5) & \multicolumn{1}{c}{---}  & 3.45(6) & 3.56(8) \\
 $^{87}$Rb-Ar &  2.79(5) & \multicolumn{1}{c}{---}  & 3.035(7) & 3.29(5) \\
 $^{87}$Rb-CO$_2$ & 2.84(6) & \multicolumn{1}{c}{---}  & \multicolumn{1}{c}{---} & \multicolumn{1}{c}{---} \\
 $^{87}$Rb-Kr &  \multicolumn{1}{c}{---} & \multicolumn{1}{c}{---} & 2.79(1) & 2.80(4) \\
 $^{87}$Rb-Xe &  2.75(4) & \multicolumn{1}{c}{---}  & 2.88(1) & 2.87(6) \\
\hline
\end{tabular}
\caption{
Comparison of this work with published  measurements, including those utilizing universality of quantum diffractive collisions (UQDC), and theoretical calculations of $^{87}$Rb-X loss rate coefficients.
Numbers in parentheses are one-standard-deviation, $k=1$ uncertainties.
For simplicity, the statistical and systematic uncertainties from Ref.~\onlinecite{Shen2021} are added in quadrature.
For the theory and this work, $T=295.2(3)$~K. For Ref.~\onlinecite{Shen2021}, $T=294$~K.
}
\label{tab:Rb_comparison}
\end{table}

\bibliography{main}